

\input phyzzx.tex
\overfullrule0pt


\def\etal{{\it et al.}}
\def\half{{\textstyle{1 \over 2}}}

\def\eighth{{\textstyle{1 \over 8}}}

\def\bold#1{\setbox0=\hbox{$#1$}%
     \kern-.025em\copy0\kern-\wd0
     \kern.05em\copy0\kern-\wd0
     \kern-.025em\raise.0433em\box0 }
\Pubnum={VAND-TH-93-13}
\date={October 1993}
\pubtype{}
\titlepage


\vskip1cm
\title{\bf The ${\bold{b\rightarrow s\gamma}}$ Decay in Supergravity
with Radiatively Electroweak Breaking}
\author{Marco Aurelio D\'\i az }
\vskip .1in
\centerline{Department of Physics and Astronomy}
\centerline{Vanderbilt University, Nashville, TN 37235}
\vskip .2in

\centerline{\bf Abstract}
\vskip .1in

It is analyzed the branching ratio $B(b\rightarrow
s\gamma)$ in the context of minimal $N=1$ supergravity with
radiatively broken electroweak symmetry group.
There is a strong dependence on supersymetric parameters, but constraints
on the charged Higgs mass in non-supersymmetric models are relaxed,
due to large
contribution from the chargino/up-type squark sector that
interacts destructively with the Standard Model and the charged Higgs
contributions. Large suppressions/enhancements of the branching fraction
are found for large values of $\tan\beta$.

\vfill

\endpage

\voffset=-0.2cm

\REF\InamiL{T. Inami and C.S. Lim, {\it Prog. Theor. Phys.} {\bf 65},
297 (1981).}
\REF\TwoHDM{T.G. Rizzo, {\it Phys. Rev. D} {\bf 38}, 820 (1988); B.
Grinstein and M.B. Wise, {\it Phys. Lett. B} {\bf 201}, 274 (1988);
W.-S. Hou and R.S. Willey, {\it Phys. Lett. B} {\bf 202}, 591 (1988);
T.D. Nguyen and G.C. Joshi, {\it Phys. Rev. D} {\bf 37}, 3220 (1988);
C.Q. Geng and J.N. Ng, {\it Phys. Rev. D} {\bf 38}, 2857 (1988);
D. Ciuchini, {\it Mod. Phys. Lett. A} {\bf 4}, 1945 (1989);
B. Grinstein, R. Springer and M. Wise, {\it Nucl. Phys.} {\bf B339},
269 (1990); V. Barger, J.L. Hewett and R.J.N. Phillips, {\it Phys.
Rev. D} {\bf 41}, 3421 (1990), and Erratum.}

The decay $b\rightarrow s\gamma$ is forbidden at tree level but
induced in the Standard Model (SM) at one loop by $W$
and Goldstone bosons together with up-type quarks in the internal
lines of the loop\refmark\InamiL. The SM value of the branching
ratio of this decay is $B(b\rightarrow s\gamma)\approx
4\times 10^{-4}$ for $m_t=140$ GeV and increases with $m_t$.
In two-Higgs-doublets models, loops involving charged Higgs bosons
and up-type quarks have to be added\refmark\TwoHDM. The contribution
from the charged Higgs boson in type II models  (where one
Higgs doublet couples to the up-type quarks and the other
Higgs doublet couples to the down-type quarks)
has the same sign as the
SM contribution. In type I models (where only one Higgs doublet couples
to the fermions) the charged Higgs boson contribution does not have a
definite sign and decreases with $\tan\beta$.
This fact is responsible for the strong constraints
on type II models in comparison with type I models.

\REF\CLEOpenguin{CLEO Collaboration, R. Ammar \etal, {\it Phys.
Rev. Lett.} {\bf 71}, 674 (1993).}
\REF\ratioBb{T. Altomari, {\it Phys. Rev. D} {\bf 37}, 677 (1988);
C.A. Dominguez, N. Paver and Riazuddin, {\it Phys. Lett. B} {\bf 214},
459 (1988); N.G. Deshpande, P. Lo and J. Trampetic, {\it Z. Phys. C}
{\bf 40}, 369 (1988); T.M. Aliev, A.A. Ovchinnikov and V.A. Slobodenyuk,
{\it Phys. Lett. B} {\bf 237}, 569 (1990); P.J. O'Donnell and H.K.K.
Tung, {\it Phys. Rev. D} {\bf 44}, 741 (1991); A. Ali and C. Greub,
{\it Phys. Lett. B} {\bf 259}, 182 (1991); A. Ali, T. Ohl and T.
Mannel, {\it Phys. Lett. B} {\bf 298}, 195 (1993).}
\REF\HewettRizzo{J.L. Hewett and T.G. Rizzo, report No. ANL-HEP-PR-93-37
(June 1993), unpublished.}
\REF\expbsf{E. Thorndike, CLEO Collaboration, talk given at the
1993 Meeting of the American Physical Society, Washington D.C.,
April 1993.}
\REF\hewettBBP{J.L. Hewett, {\it Phys. Rev. Lett.} {\bf 70}, 1045
(1993); V. Barger, M.S. Berger and R.J.N. Phillips, {\it Phys. Rev.
Lett.} {\bf 70}, 1368 (1993).}
\REF\misiak{M. Misiak, {\it Nucl. Phys.} {\bf B393}, 23 (1993).}
\REF\diazbtosf{M.A. D\'\i az, {\it Phys. Lett. B} {\bf 304},
278 (1993).}
\REF\ChargedH{J.F. Gunion and A. Turski, {\it Phys. Rev. D} {\bf 39},
2701 (1989);{\bf 40}, 2333 (1989); A. Brignole, J. Ellis, G. Ridolfi
and F. Zwirner, {\it Phys. Lett. B} {\bf 271}, 123 (1991); M. Drees
and M.M. Nojiri, {\it Phys. Rev. D} {\bf 45}, 2482 (1992);
A. Brignole, {\it Phys. Lett. B} {\bf 277}, 313 (1992);
P.H. Chankowski, S. Pokorski and J. Rosiek,
{\it Phys. Lett. B} {\bf 274}, 191 (1992);
M.A. D\'\i az and H.E. Haber, {\it Phys. Rev. D}
{\bf 45}, 4246 (1992).}
\REF\diaz{M.A. D\'\i az, {\it Phys. Rev. D} {\bf 48}, 2152 (1993);
M.A. D\'\i az, {\it The Fermilab Meeting DPF'92}, ed. by C. Albright
\etal, World Scientific, page 1194.}

The first observation of the exclusive decays $B^0\rightarrow K^*
(892)^0\gamma$ and $B^-\rightarrow K^*(892)^-\gamma$ by the CLEO
Collaboration\refmark\CLEOpenguin is a strong evidence for the penguin
diagrams at the quark level process $b\rightarrow s\gamma$.
The ratio between the exclusive and the inclusive decays
$\Gamma(B\rightarrow K^*\gamma)/\Gamma(b\rightarrow s\gamma)$
has been calculated\refmark\ratioBb, but with a high uncertainty:
the predicted values range from 4\% and 40\%. This impose a lower
bound on the inclusive branching ratio whose conservative estimation
\refmark\HewettRizzo is $B(b\rightarrow s\gamma)>0.65\times 10^{-4}$.
The latest experimental upper bound on the branching fraction
for the inclusive decay mode $b\rightarrow s\gamma$, given by
$B(b\rightarrow s\gamma)<5.4\times 10^{-4}$ at 90\% c.l.
\refmark\expbsf, sets powerful constraints on the charged
Higgs boson mass in two Higgs doublets models of type II
\refmark{\hewettBBP}. Other corrections
have been calculated recently: next-to-leading
logarithmic QCD-corrections\refmark\misiak, and electroweak
corrections in the context of supersymmetry\refmark\diazbtosf
to the charged Higgs mass\refmark\ChargedH and to the charged
Higgs-fermion-fermion vertex\refmark\diaz.

\REF\BBMR{S. Bertolini, F. Borzumati, A. Masiero and G. Ridolfi,
{\it Nucl. Phys.} {\bf B353}, 591 (1991).}
\REF\susy{S. Bertolini, F. Borzumati and A. Masiero, {\it Nucl. Phys.}
{\bf B294}, 321 (1987); S. Bertolini, F.M. Borzumati, and A. Masiero,
{\it Phys. Lett. B} {\bf 192}, 437 (1987); T.M. Aliev and M.I.
Dobroliubov, {\it Phys. Lett. B} {\bf 237}, 573 (1990).}
\REF\Barbieri{R. Barbieri and G.F. Giudice, {\it Phys. Lett. B}
{\bf 309}, 86 (1993).}
\REF\masSUSY{N. Oshimo, {\it Nucl. Phys.} {\bf B404}, 20 (1993);
J.L. Lopez, D.V. Nanopoulos, and G.T. Park,
{\it Phys. Rev. D} {\bf 48}, 974 (1993);
Y. Okada, {\it Phys. Lett. B} {\bf 315}, 119 (1993);
R. Garisto and J.N. Ng, {\it Phys. Lett. B} {\bf 315}, 372 (1993).}
\REF\guide{J.F. Gunion, H.E. Haber, G. Kane and S. Dawson,
{\it The Higgs Hunter's Guide} (Addison-Wesley, Reading MA, 1990).}

In supersymmetry, the contributions from charginos together with
up-type squarks and from neutralinos and gluinos together with
down-type squarks, have to be included\refmark{\BBMR,\susy}.
It was stressed that in this case, it is important
the effect of loops involving
supersymmetric particles\refmark{\Barbieri,\masSUSY}. Here
we study this effect in the context of the radiatively
broken Minimal Supersymmetric Model\refmark\guide,
following ref. [\BBMR] and including the effects
described in refs. [\misiak,\diazbtosf].

\REF\susyrep{H.P. Nilles, {\it Phys. Rep} {\bf 110}, 1 (1984);
H.E. Haber and G.L. Kane, {\it Phys. Rep.} {\bf 117}, 75 (1985).}

Minimal $N=1$ supergravity\refmark\susyrep is characterized
by the superpotential
$$W=h_U^{ij}Q_iU_j^cH_2+h_D^{ij}Q_iD_j^cH_1+h_E^{ij}L_iE_j^cH_1
+\mu\varepsilon_{ab}H_1^aH_2^b\eqn\suppot$$
where $i,j=1,2,3$ are indices in generation space, $\varepsilon_{ab}$
with $a,b=1,2$ is the antisymmetric tensor in two dimensions,
and $\mu$ is the Higgs mass parameter.
The $3\times 3$ matrices $h_U, h_D$ and $h_E$ are the Yukawa couplings.
The soft supersymmetry breaking terms are
$${\cal L}_s=A_U^{ij}h_U^{ij}\tilde Q_i\tilde U^c_jH_2+A_D^{ij}
h_D^{ij}\tilde Q_i\tilde D_j^cH_1+A_E^{ij}h_E^{ij}\tilde L_i
\tilde E_j^cH_1+B\mu\varepsilon_{ab}H_1^aH_2^b+h.c.\eqn\softab$$
plus a set of scalar and gaugino mass terms, which at the unification
scale are
$${\cal L}_m=m_0^2\sum_i|S_i|^2+\big[\half M_{1/2}(\lambda_1
\lambda_1+\lambda_2\lambda_2+\lambda_3\lambda_3)+h.c.\big]
\eqn\softmass$$
where $S_i$ are all the scalars of the theory and $\lambda_i, i=1,2,3$
are the gauginos corresponding to the groups $U(1), SU(2)$ and $SU(3)$
respectively. In eq.~\softab\ all the fields are scalar components of
the respective superfields.
The mass parameters $A$ and $B$ are of ${\cal O}(m_0)$ and in some
supergravity models they satisfy the following relation at the
unification scale $M_X$: $A=B+m_0$
where $A$ is the common value for the $A_a^{ii}$ ($a=U,D,E$) parameters
at $M_X$: $A_U^{ij}=A_D^{ij}=A_E^{ij}=A\delta^{ij}$.

\REF\KLNPY{S. Kelley, J. Lopez, D. Nanopoulos, H. Pois and K. Yuan,
{\it Nucl. Phys.} {\bf B398}, 3 (1993).}
\REF\diazhaberii{M.A. D\'\i az and H.E. Haber,
{\it Phys. Rev. D} {\bf 46}, 3086 (1992).}

At the weak scale, the tree level Higgs potential is given by
$$\eqalign{V=&m_{1H}^2|H_1|^2+m_{2H}^2|H_2|^2-m_{12}^2(H_1H_2+h.c.)
\cr&+\eighth(g^2+g'^2)(|H_1|^2-|H_2|^2)^2+\half g^2|H_1^*H_2|^2
\cr}\eqn\Vtree$$
where $m_{iH}^2=m_i^2+|\mu|^2$ ($i=1,2$) and
$m_{12}^2=-B\mu$. The two Higgs doublets mass parameters $m_1$ and
$m_2$ satisfy $m_1=m_2=m_0$ at the unification scale.
The three mass parameters in eq.~\Vtree\ can be replaced by the $Z$
boson mass $m_Z$, the CP-odd Higgs mass $m_A$, and the ratio between
the vacuum expectation values of the two Higgs doublets $\tan\beta
\equiv v_2/v_1$, according to the formulas
$$\eqalign{m_{1H}^2=&-\half m_Z^2c_{2\beta}+\half m_A^2(1-c_{2\beta})
\cr m_{2H}^2=&\half m_Z^2c_{2\beta}+\half m_A^2(1+c_{2\beta})\cr
m_{12}^2=&\half m_A^2s_{2\beta}\cr}\eqn\conversion$$
where $s_{2\beta}$ and $c_{2\beta}$ are sine and cosine functions of the
angle $2\beta$. The previous relations are valid at tree level. The
effects of the one-loop corrected Higgs potential may be important in
some cases\refmark\KLNPY, especially near $\tan\beta=1$ when $m_{1H}=
m_{2H}=m_{12}$ and the lightest neutral Higgs mass comes only from
radiative corrections\refmark\diazhaberii.

\REF\DreesNojiri{M. Drees and M.M. Nojiri, {\it Nucl. Phys.}
{\bf B369}, 54 (1992).}
\REF\diazargonne{M.A. D\'\i az, report No. VAND-TH-93-11, presented at
the Workshop on Physics at Current Accelerators and the Supercollider,
Argonne National Laboratory, Argonne Il, June 2-5 1993.}
\REF\rgesol{G.G. Ross and R.G. Roberts, {\it Nucl. Phys.} {\bf B377},
571 (1992).}
\REF\falck{N.K. Falck, {\it Z. Phys. C} {\bf 30}, 247 (1986).}

The electroweak symmetry group is broken radiatively when the different
parameters are evolved from the grand unification scale to the weak scale
\refmark\DreesNojiri.
In ref.~[\diazargonne] can be found a typical solution of the
renormalization
group equations (RGE) in the spirit of ref.~[\rgesol], but including the
trilinear $A$ parameters and other Higgs mass parameters as well. The
effects of the supersymmetric threshold are neglected. The RGE
used are given in ref.~[\BBMR]\ with the
exception of the $A$ parameters, whose equations are taken from
ref.~[\falck]. The set of independent parameters is chosen to be $m_t,
m_A$ and $\tan\beta$ at the weak scale, $M_{1/2}$ at the unification
scale, and the sign of $\mu$ as a discrete parameter.

\REF\QCDcorr{S. Bertolini, F. Borzumati and A. Masiero, {\it Phys. Rev.
Lett.} {\bf 59}, 180 (1987); N.G. Deshpande, P. Lo, J. Trampetic,
G. Eilam and P. Singer, {\it Phys. Rev. Lett.} {\bf 59}, 183 (1987);
B. Grinstein, R. Springer and M.B. Wise, {\it Nucl. Phys.} {\bf
B339}, 269 (1990); P. Cho and B. Grinstein, {\it Nucl. Phys.}
{\bf B365}, 279 (1991); M. Misiak, {\it Phys. Lett. B} {\bf 269},
161 (1991); A.J. Buras, M. Jamin, M.E. Lautenbacher and P.H.Weisz,
{\it Nucl. Phys.} {\bf B370}, 69 (1992); M. Misiak, {\it Nucl. Phys.}
{\bf B393}, 23 (1993); K. Adel and Y.-P. Yao, report No.
UM-TH-93-20 (August 1993), unpublished.}

The QCD uncorrected amplitude for the decays $b\rightarrow s\gamma$
and $b\rightarrow sg$ are
$$A^{\gamma,g}(m_W)=A_{SM}^{\gamma,g}+(f^+f^-)
A_{H^{\pm}}^{1\gamma,g}+(f^-)^2\cot^2\beta A_{H^{\pm}}^{2\gamma,g}+
A_{\tilde\chi^{\pm}}^{\gamma,g}+A_{\tilde g}^{\gamma,g}
\eqn\Atotfg$$
where the form factors $f^{\pm}$ come from the renormalization
of the charged Higgs boson coupling to a pair of fermions\refmark\diaz.
The different amplitudes $A$ can be found in ref.~[\BBMR].
If we now run the scale from $m_W$ to $m_b$ and introduce the QCD
corrections\refmark\QCDcorr, we get
$$A^{\gamma}(m_b)=\eta^{-{\textstyle{16\over 23}}}\Bigg[
A^{\gamma}(m_W)+{8\over 3}A^g(m_W)(\eta^{\textstyle{2\over 23}}-1)
\Bigg]+CA_0^{\gamma}\eqn\AcorrQCD$$
where $\eta=\alpha_s(m_b)/\alpha_s(m_W)\approx 1.83$ and $A_0^{\gamma}$
is given by
$$A_0^{\gamma}={{\alpha_W\sqrt{\alpha}}\over{2\sqrt{\pi}}}{{V_{ts}^*
V_{tb}}\over{m_W^2}}\eqn\azero$$
with $C=0.177$, $\alpha_W=g^2/4\pi$, and $\alpha=e^2/4\pi$. This last
term proportional to
$C$ comes from mixing of four quark operators\refmark\Barbieri.

\FIG\bsfi{Branching ratio of the inclusive decay $B(b\rightarrow
s\gamma)$ as a function of $\tan\beta$ for the SM and for the
SUSY-GUT model. It is also displayed the CLEO upper bound.}

In fig.~\bsfi\ it is plotted the branching ratio of the inclusive decay
$B(b\rightarrow s\gamma)$ as a function of $\tan\beta$. In this model
it is not possible to get the correct electroweak symmetry breaking
if $\tan\beta\gsim m_t/m_b$. If $\tan\beta$ increases
the scalar mass parameter $m_0$ must grow
with $\tan\beta$ in order to get the necessary splitting $m_{2H}^2-
m_{1H}^2$, producing heavy squarks and consequently a suppressed
contribution to the branching ratio from the chargino/up-type squarks
loops. This in turn produce a large value for $B(b\rightarrow s
\gamma)$. On the other side, $\tan\beta$ is bounded from below by
$\tan\beta\gsim 1$, since it is not possible neither to get the correct
electroweak symmetry breaking otherwise.

\FIG\bsfii{Same than fig.~\bsfi\ but as a function of $m_A$.}

In fig.~\bsfii\ we see the dependence of $B(b\rightarrow s\gamma)$
on the mass of the CP-odd Higgs mass $m_A$. For large values of $m_A$
the contribution of the charged Higgs mass decreases since its mass
increases. But also the scalar mass parameter $m_0$ increases with
$m_A$, thus making heavy squarks while keeping almost constant
chargino and neutralino masses. This makes the magnitude of the chargino
contribution (it has a negative sign with respect to the $W$ and
$H^{\pm}$ contributions) to $B(b\rightarrow s\gamma)$ decrease
also with $m_A$ but faster than the charge Higgs contribution. The net
effect is a growing $B(b\rightarrow s\gamma)$ with $m_A$, and
approaching to the SM value.

\FIG\bsfiii{Same than fig.~\bsfi\ but as a function of $m_t$.}

The dependence on the
top quark mass of the branching ratio $B(b\rightarrow s\gamma)$ can be
seen in fig.~\bsfiii. In the SM this branching ratio grows with the top
quark mass and remains below the CLEO bound in the hole range of $m_t$.
In the SUSY--GUT model, for the parameters considered here, the branching
ratio decreases with the top quark mass except for very large values of
$m_t$. This effect is due to a faster growing (and with opposite sign)
chargino contribution compared to the charged Higgs contribution.
This can be seen also in ref.~[\Barbieri]
taking into account that in the SUSY--GUT model, a change in $m_t$ implies
a change in $m_0$, in opposition to the treatment in
ref.~[\Barbieri] where both parameters are independent. If the top quark
mass increases, the splitting $m_1^2-m_2^2$ at the weak scale increases
also, and to keep it constant $m_0$ must decrease. This in turn will
decrease the absolute values of $m_1^2$ and $m_2^2$, so $\mu$ will grow
in order to keep $m_{1H}^2$ and $m_{2H}^2$ unchanged.

\FIG\bsfiv{Same than fig.~\bsfi\ but as a function of $M_{1/2}$.}

In fig.~\bsfiv\ it is plotted the branching ratio $B(b\rightarrow
s\gamma)$ as a function of the gaugino mass parameter $M_{1/2}$ (at the
unification scale $M_X$). For small values of $M_{1/2}$ the mass of the
lightest chargino and neutralino become too small. The relatively
weak dependence of $B(b\rightarrow s\gamma)$ on $M_{1/2}$ appears
because the masses of the charginos and up-type squarks grows slowly
with $M_{1/2}$ and at a comparable rate.

We have studied the prediction for the branching ratio $B(b\rightarrow
s\gamma)$ in the context of minimal $N=1$ supergravity with a radiatively
broken electroweak symmetry group. We found convenient to parametrize
the model with the following independent parameters: $m_t$, $m_A$ and
$\tan\beta$ at the weak scale, $M_{1/2}$ at the unification scale, and
the sign of $\mu$ as a discrete parameter.
It is clear that the experimental upper bound of the branching ratio
of the decay $b\rightarrow s\gamma$ does not strongly constrain the
charged Higgs mass, as it does to non-supersymmetric two-Higgs-doublet
models of type II. The branching ratio $B(b\rightarrow s\gamma)$
lies below the CLEO bound and even below the
SM value in some regions of the parameter space explored here, although
the opposite also occurs. There
is an important dependence on $\tan\beta$, and
large suppressions/enhancements of $B(b\rightarrow s\gamma)$
with respect to the Standard Model are
possible for large values of this parameter. We have include in the
calculation QCD corrections and some electroweak radiative corrections
as well. The later are more important at large values of $\tan\beta$.

Given the strong dependence of $B(b\rightarrow s\gamma)$ on the
different supersymmetric parameters it is possible to rule out some
regions of the parameter space and, because of this reason, we emphasize
that an experimental measurement of this decay, or an improvement on the
upper bound, will be an important way to test supersymmetry.

\REF\BorzuLopez{J.L. Lopez, D.V. Nanopoulos, G.T. Park and A. Zichichi,
report No. CERN-TH.6979/93, July 1993, unpublished;
F.M. Borzumati, report No. DESY 93-90, August 1993, unpublished.}

\vskip .5cm
\noindent{\bf Note added:}
When this work was completed, we received two
preprints\refmark\BorzuLopez
that calculate the branching ratio $B(b\rightarrow s\gamma)$ in the MSSM
with a radiatively induced breaking of the gauge group $SU(2)\times
U(1)$. In the DESY preprint, a strong dependence on $\tan\beta$ is also
observed when this parameter is large. In the CERN preprint, although
the models considered have a more restricted parameter space than ours,
the dependence of the
branching ratio on $\tan\beta$ is similar.
\vskip .5cm

\vskip .5cm
\centerline{\bf ACKNOWLEDGMENTS}
\vskip .5cm

Discussions with Howard Baer, Joseph Lykken, Xerxes Tata, and Thomas
Weiler are gratefully acknowledged.
I am thankful to the Particle Theory Group at Fermilab, where part of
this work was completed.
This work was supported by the U.S. Department of Energy, grant No.
DE-FG05-8SER40226.

\refout
\figout
\end